\pdfoutput=1 
\documentclass[11pt, letterpaper]{article}
\usepackage{mydef2col}
\usepackage{kantlipsum,widetext}
\usepackage[utf8]{inputenc}
\usepackage[T1]{fontenc}
\usepackage[autostyle=true]{csquotes}
\usepackage{regexpatch}
\usepackage{cprotect}

\allowdisplaybreaks[4]
\graphicspath{{./Graphs/}}
\usepackage[colorinlistoftodos,prependcaption,textsize=tiny]{todonotes}
\makeatletter
\xpatchcmd{\@todo}{\setkeys{todonotes}{#1}}{\setkeys{todonotes}{inline,#1}}{}{}
\makeatother

\usepackage[square]{natbib}

\def\calF{{\cal F}}
\def\calS{{\cal S}}
\def\calV{{\cal V}}

\def\calVH{{\cal V}_H}
\def\calY{{\cal Y}_H}

\def\Ito{It\^{o}'s }

\def\bT{{\bar{T}}}
\def\pade{Pad{\'e} }
\def\Imm{\operatorname{Im}}
\def\Ree{\operatorname{Re}}

\newcommand{\iu}{\mathrm{i}\mkern1mu}

\title{The ATM implied skew in the ADO-Heston model}

\author{
\authorstyle{Andrey Itkin}
\newline\newline
\institution{Tandon School of Engineering, New York University, USA.}
}

\date{\today}

\begin{document}

\maketitle

\lettrineabstract{In this paper similar to [P. Carr, A. Itkin, 2019] we construct another Markovian approximation of the rough Heston-like volatility model - the ADO-Heston model. The characteristic function (CF) of the model is derived under both risk-neutral and real measures which is an unsteady three-dimensional PDE with some coefficients being functions of the time $t$ and the Hurst exponent $H$. To replicate known behavior of the market implied skew we proceed with a wise choice of the market price of risk, and then find a closed form expression for the CF of the log-price and the ATM implied skew. Based on the provided example, we claim that the ADO-Heston model (which is a pure diffusion model but with a stochastic mean-reversion speed of the variance process, or a Markovian approximation of the rough Heston model) is able (approximately) to reproduce the known behavior of the vanilla implied skew at small $T$. We conclude that the behavior of our implied volatility skew curve ${\cal S}(T) \propto a(H) T^{b\cdot (H-1/2)},  \, b = const$, is not exactly same as in rough volatility models since $b \ne 1$, but seems to be close enough for all practical values of $T$. Thus, the proposed Markovian model is able to replicate some properties of the corresponding rough volatility model. Similar analysis is provided for the forward starting options where we found that the ATM implied skew for the forward starting options can blow-up for any $s > t$ when $T \to s$. This result, however, contradicts to the observation of [E. Alos, D.G. Lorite, 2021] that Markovian approximation is not able to catch this behavior, so remains the question on which one is closer to reality.
}

\vspace{0.5in}

\section*{Introduction}

Rough volatility models have been acquiring an increasing popularity since it was shown in \citep{GatheralJaissonRos2014} that for a wide range of assets, historical volatility time-series exhibit a behavior which is much rougher than that of the Brownian motion (BM)\footnote{Some theoretical foundation for this approach was built even earlier, e.g., in \citep{ALos2007}.}.  Nowadays, there exists a vast literature on the subject which on a permanent basis is collected in \citep{RVsite}. One of the important findings of rough volatility models consists in their ability to reproduce the explosive behavior of the implied at-the-money (ATM) skew observed empirically when the option maturity goes to zero, see \citep{Bayer2016, Fukasawa2011} and a more extended surveys in \citep{AlosBook, AlosLeon2021} among others.

The cost one has to pay for getting all these advantages of the rough volatility models are technical problems arising due to a non-Markovian nature of the fractional BM (fBM). Therefore, e.g., derivatives pricing becomes more complicated than in standard stochastic volatility models (SVM), despite there exists a significant progress on this way as well. For instance,  in \citep{EuchRos2016} using an original link between nearly unstable Hawkes processes and fractional volatility models, the authors derived a semi-closed formula for the characteristic function in the rough Heston model. The formula is not fully explicit but given in terms of the solution of a fractional Riccati equation. To avoid the computational burden arising from the numerical solution of this fractional Riccati equation, in \citep{Baschetti2021} the \pade approximant of the solution was used which has been discussed in \citep{GatheralRadoicic2019}. The authors show that the rational approximation provides a very accurate description of the solution, especially for low values of the Hurst exponent $H$. They also claim results of empirical investigations, both under the pricing and the historical measures, which reveal $H$ to be of order 0.05-0.1, thus motivating the use of the rational approximation.

Pricing variance swaps is even more complicated, see e.g., \citep{JacquierMuguruzaPannier2021} and references therein. Therefore, various Markovian approximation to the fBM were proposed to simplify pricing and hedging of derivatives securities which are able to reproduce some properties of the rough volatility models. For instance, in \citep{Muravlev2011} it is shown that the fBM can be represented as a linear functional of an infinite-dimensional Markov process. Later in \citep{Harms2021} the fBM was represented as an integral over a family of the Ornstein-Uhlenbeck (OU) processes. The author proposes numerical discretizations which have strong convergence rates of an arbitrarily high polynomial order. He uses this representation as the basis for constructing some Monte Carlo schemes for fractional volatility models, e.g., the rough Bergomi model.

In \citep{Rogers2019} a simpler alternative to rough volatility is proposed which is represented by a mixture of two correlated OU processes. It is based on empirical observation of daily volatility level estimates for the S\&P 500 index: the level fluctuates strongly on small time scales, but on longer time scales it seems to be changing. Therefore, if the level did not change, the data is modeled by the OU process with strong mean reversion and high volatility; as the level appears to be changing, it is modeled by an energetic OU process mean-reverting to a slower one. The author concludes that his OU-OU model on timescales of days, weeks and months works well (at least, not worse than the rough volatility models), and is much easier to deal with being a bivariate Gaussian diffusion, amenable to the multiscale option pricing techniques. It is also worth mentioning that the approach of \citep{Rogers2019} can be considered as a simplistic approximation of the approach in \citep{Harms2021} where only two OU processes out of the whole family are taken into account.

Another type of approximations was proposed in \citep{Bayer2021}. The authors consider rough SVM where the variance process $v_t$ satisfies a stochastic Volterra equation with a fractional kernel, as in the rough Bergomi and the rough Heston models and claim that simulation of such rough processes often results in high computational cost. That is why they propose approximations of stochastic Volterra equations using an $N$-dimensional diffusion process defined as solution to a system of ordinary stochastic differential equation and show that under some regularity conditions these approximations converge strongly with a super-polynomial rate in $N$.

In \citep{ADOL_Risk} another Markovian approximation of the fBM, known as the Dobric-Ojeda (DO) process, was applied to the fractional SVM where the instantaneous volatility $\sigma_t$ is modeled by a lognormal process with drift and fractional diffusion. Since the DO process is a semi-martingale, it can be represented as an \Ito diffusion. It turns  out that in this framework (the ADOL model):
\begin{itemize}
\item the process for the spot price $S_t$ is a geometric BM with the stochastic instantaneous volatility $\sigma_t$;
\item the process for $\sigma_t$ is also a geometric BM with the stochastic speed of mean reversion and time-dependent volatility of volatility;
\item the supplementary process $\calV_t$ is the OU process with time-dependent coefficients and is also a function of the Hurst exponent.
\end{itemize}
The authors also introduce an adjusted DO process which provides a uniformly good approximation of the fBM for all Hurst exponents $H \in [0,1]$ but requires a complex measure. Finally, the characteristic function (CF) of $\log S_t$ in this model can be found in closed form by using asymptotic expansion. Therefore, pricing options and variance swaps (by using a forward CF) can be done via fast Fourier transform (FFT), which is much easier than  in rough volatility models. It can be seen that the DO process can also be considered as a particular case of the construction in \citep{Harms2021}, however, providing some additional tractability, while, perhaps, is less accurate.

A quite different approach  has been utilized in \citep{FrizPigatoSeibel2021} who proposed the Step Stochastic Volatility Model (SSVM).  The authors were looking for a (possible) simple (without adding jumps or non-Markovian rough fractional volatility dynamics) modification of a class of SVMs to be capable of producing extreme short-dated implied volatility skew. Indeed, much of the recent success of the rough SVMs is due to the fact that they are capable to predict a desirable (observed at the market) behavior of the ATM implied skew. As shown in \citep{Fukasawa2011, Bayer2018} among others, the ATM implied skew blows up at the rate $T^{H-1/2}$ when $T \to 0$, where $T$ is the time to maturity, and the Hurst parameter $H \in (0, 1/2]$ quantifies the roughness of the volatility process. Accordingly, the blowup can be at most of the order $T^{-1/2}$ which is a model-free consequence of no-arbitrage, \citep{Lee2002}. But in \citep{FrizPigatoSeibel2021} the authors introduced a leverage effect by making volatility discontinuous at the money. To achieve this goal, they multiplied the (backbone) stochastic volatility with distinct factors, say $\sigma_-$ and $\sigma_+$ depending on whether the considered option is out-of-the-money or in-the-money), and the implied skew generated by such a model explodes as $T^{-1/2}$, \citep{Pigato2019}. However, this model is not able to predict the short end implied ATM skew blow up at the rate $T^{H-1/2}$.

Also, based on the relationship between forward skews and vanilla skews derived in \citep{AlosBook} by using Malliavin calculus, the authors show that models constructed based on the fBM are able to reproduce the blow-up of the forward skew, but those based on the Markovian approximation miss this feature. The reason is that the vanilla skew can blow up at every time $s > t$ for fractional volatilities, while not for the Markovian approximations, that blows up only at $t=0$. And this result is model independent, i.e., valid for any diffusive model. See also a recent paper \citep{Alos2022}.

Therefore, in this paper we are interested in an independent check of this result and aim to closely look at the behavior of the implied skew in the Markovian approximations of the rough volatility models. For doing that we use the approach of \citep{ADOL_Risk}, but for better tractability use a modified version of the ADO model where the stochastic variance (rather the volatility in \citep{ADOL_Risk}) follows the CIR process (like in the Heston model) combined with the ADO construction. In what follows we call this model as the ADO-Heston model.

At the end of this introduction, we have to mention two recent papers on the subject which do question market trends that stimulated the original development of the rough volatility models. In particular, in  \citep{ContRough2022} the authors used a model-free approach and analysed statistical evidence for the use of rough fractional processes with Hurst exponent $H<0.5$. For doing so they introduce a non-parametric method for estimating the roughness of a function based on a discrete sample. They further investigate the finite sample performance of this estimator for measuring the roughness of sample paths of stochastic processes. The authors also describe detailed numerical experiments provided based on sample paths of fractional Brownian motion and other fractional processes. Based on the results obtained, they claim that those numerical experiments based on stochastic volatility models show that, even when the instantaneous volatility has a diffusive dynamics with the same roughness as the Brownian motion, the realized volatility exhibits rough behavior corresponding to the Hurst exponent significantly smaller than 0.5. The conclusion is made that irrespective of the roughness of the spot volatility process, the realized volatility always exhibits rough behavior with an apparent Hurst index $H<0.5$.

The second paper, \citep{Gyuon2022}  analyses the term structure of the ATM skew of equity indexes by using two years data of S\&P~500, Eurostoxx 50, and DAX. The authors find that this skew does not follow a power law for short maturities and is better captured by simple parameterizations that do not blow up for vanishing maturity. For instance, the ATM skew produced by the two-factor Bergomi model provides the best fits. Two other models built using non-blowing-up kernels are introduced in this paper and also demonstrate similar results. In contrast, the fits of the rough Bergomi model and power law deteriorate quickly as the time gets closer to the first monthly options maturity. The extrapolated zero-maturity skew is far from being infinite and is distributed around the point 1.5 (in absolute value).

From the modeling point of view these results mean that, perhaps, market data on realized volatility are not sufficient to decide which rough or Markovian stochastic volatility model is preferable to replicate the observed market behavior. Therefore, other measures would be useful for this purpose, e.g., the vanilla and forward implied volatilities and skews which could be retrieved from the market data. For the reference, again see \citep{AlosBook}.

The rest of the paper is organized as follows. In Section~\ref{ADOH} we briefly describe our modification of the ADOL model - the ADO-Heston model. In Section~\ref{ADOHPDE} an ADO-Heston PDE for the option price is derived under both risk-neutral and real measures which is an unsteady three-dimensional PDE with some coefficients being functions of the time $t$ and $H$. In Section~\ref{CFT} we further specify the model by choosing a special form of the market price of risk function, and then find a closed form expression for the CF of the log-price and the ATM implied skew. Using an example described in Section~\ref{example}, we claim that the ADO-Heston model (which is a pure diffusion model but with a stochastic mean-reversion speed of the variance process, or a Markovian approximation of the rough Heston model) is able (approximately) to reproduce the known behavior of the vanilla implied skew at small $T$. Section~\ref{disc} provides a preliminary discussion of the results obtained. We conclude that the behavior of our implied volatility skew curve
${\cal S}(T) \propto a(H) T^{b\cdot (H-1/2)},  \, b = const$, is not exactly same as in rough volatility models since $b \ne 1$, but seems to be close enough for all practical values of $T$. Thus, the proposed Markovian model is able to replicate some properties of the corresponding rough volatility model. Section~\ref{fso} provides a similar analysis for the forward started options. The final section concludes.

\section{The ADO-Heston model} \label{ADOH}

Among various SV models the classical Heston model, \citep{Heston:93}, is one of the most popular mainly due to its tractability. In \citep{GueJacRoome2014,EuchRos2016} fractional versions of the Heston model were proposed with the Hurst parameter $H \in [0, 1/2]$. By construction the underlying process is neither Markovian, nor a semi-martingale. It was discovered that the characteristic function of the log-price in rough Heston models exhibits the same structure as that one in the classical Heston model but with the Riccati equation replaced by its fractional version. This  equation doesn't have an explicit  solution anymore but can be solved numerically by transforming it to some Volterra equation. Other rough volatility models experience similar problems while nowadays many of them are efficiently treated numerically, \citep{RVsite}.

To improve tractability and simplify this model while trying to keep its main properties, we make two steps.

\paragraph{The ADO process.} First, following the idea of \citep{ADOL_Risk}, we replace the instantaneous variance process (the fBM) with the ADO (adjusted Dobric-Ojeda) process, which is described in detail in \citep{ADOL_Risk}. In short, the Dobric-Ojeda (DO) process $V_H(t), \ t \in [0,\infty], \ H \in [0,1]$ was invented in \citep{DobricOjeda2009} as a Gaussian Markov process with similar properties to those of the fBM, namely: its increments are dependent in time. The DO process is defined by first considering the fractional Gaussian field $Z = Z_H(t), \  (t,H)\in [0,\infty)\times (0,1)$ on a probability space $(\Omega, \cal F, \mathbb{P})$ defined by covariance (compare this with a standard fBM where $\alpha_{H,H'} = 1$, and $H = H'$)
\begin{widetext}
\begin{align}
\EE [Z_H(t) Z_{H'(t)}] &= \frac{\alpha_{H,H'}}{2}\left[|t|^{H+H'} + |s|^{H+H'} - |t-s|^{H+H'}\right], \\
\alpha_{H,H'} &=
\begin{cases}
-\dfrac{2 \eta}{\pi} \xi(H) \xi(H') \cos \left[\dfrac{\pi}{2}(H'-H)\right]
\cos \left[\dfrac{\pi}{2}(H'+H)\right], & H=H' \ne 1, \\
\xi \sin^2(\pi H) \equiv \alpha_h \equiv \alpha_{H'}, & H + H' = 1,
\end{cases}
\nonumber \\
\xi(H) &= \left[\Gamma(2H+1)\sin(\pi H)\right]^{1/2}, \quad
\eta = \Gamma(-(H+H')), \quad
\xi = \left[\Gamma(2H+1)\Gamma(3-2H)\right]^{1/2}. \nonumber
\end{align}
\end{widetext}
\noindent Here $\Gamma(x)$ is the Gamma function, \citep{as64}. Obviously, if $H=H'$, $Z_H$ is a fBM, and so if $H=H'=1/2$ it is a standard BM. It was established in \citep{DObook2006} that $Z_H$ exists.

Further \citep{DobricOjeda2009} are seeking for a process of the form $\psi_H(t) M_H(t)$ that in some sense approximates fBM, assuming that $\psi_H(t)$ is a deterministic function of time, and $M_H(t)$ is a stochastic process. Omitting the details, this construction yields
\begin{align} \label{psi}
\psi_H(t) &= \dfrac{\Gamma(3-2H)}{c_H \Gamma^2(3/2-H)} t^{2H-1}, \\
c_H &=  \dfrac{\alpha_H}{2 H \Gamma(3/2-H) \Gamma(H+1/2)}, \nonumber
\end{align}
\noindent and
\begin{equation} \label{MH}
M_H(t) = \EE[Z_{H'}(t) | \calF_t^H],
\end{equation}
\noindent where $\calF_t^H$ is a filtration generated by a sigma-algebra $Z_H(s)), \ 0 \le s \le t$. It is proved in \citep{Conus2016,DobricOjeda2009}, that $M_H(t)$ is a martingale with respect to $(\calF_t^H)_{t \ge 0}$. It is also shown that $M_H(t)$ is  a Gaussian centered process with independent increments and covariance
\begin{align} \label{EMM}
\EE [M_H(t) M_H(s)] &= c_H \alpha_H \bar{B}(3/2-H)(s \wedge t)^{2-2 H}, \qquad
c_H =  \dfrac{\alpha_H}{2 H \Gamma(3/2-H) \Gamma(H+1/2)},
\end{align}
\noindent where $\bar{B}(x) = B(x,x), \ B(x,y)$ is the Beta function, and $H+H'=1$.

The most useful property of the DO process is that it is a semi-martingale and can be represented as an \Ito diffusion. This means, see again \citep{DobricOjeda2009, Wildman2016}, that there exists a BM process $W_t, \  t \in [0,\infty)$ adapted to the filtration $\calF_t^H$, such that
\begin{align} \label{DifRepr}
dV_H(t) &= \dfrac{2H-1}{t} V_H(t) dt + B_H t^{H-1/2} d W_t, \qquad
B_H = \frac{2^{3-4 H} \csc ^4(\pi  H) \Gamma (2-H)}{\Gamma \left(3/2-H\right)^2 \Gamma (H)}.
\end{align}

As shown in \citep{Guli2023}, although the factor $1/t$ in the drift of $V_H(t)$ in the equation in \eqref{DifRepr} is singular at $t=0$, the DO process does not explode at $t=0$. Indeed, using \eqref{psi} and \eqref{EMM}, we obtain $\mathbb{E}\left[V_H(t)^2\right] = C_H t^{2H}$. Hence, $\mathbb{E}\left[V_H(0)^2\right] = 0$ and, therefore,
$V_H(0) = 0$ $\mbox{a.s.}$ It follows that the natural initial condition for the DO-process is zero. Moreover, the \eqref{DifRepr} exists, and hence the $DO$-process is an \Ito diffusion. The previous statement can be derived by analysing the equation in \eqref{DifRepr} and using the following two facts
\begin{itemize}
\item The function $t\mapsto t^{H-\frac{1}{2}}$ is square-integrable over $[0,T]$. \\

\item $\int_0^T\mathbb{E}\frac{\left[|V_H(t)|\right]}{t}dt < \infty$.
\end{itemize}

The inequality in the second item can be easily derived using H\"{o}lder's inequality and \eqref{EMM}. \hfill $\square$

Note that a proof of the fact that the equation in \eqref{DifRepr} is well-defined can be found in \citep{Conus2016}, Proposition 2.5. Conus and Wildman also suggested to modify the DO  process. For every $\varepsilon > 0$, they defined a new process $t \mapsto V^{\varepsilon}_t$ in which the drift is zero until $t = \varepsilon$. As was mentioned above, this modification is not needed to account for the singularity in \eqref{DifRepr}. In \citep{Conus2016}, the modification is used to introduce a modified model and study risk-free measures in it.

As shown in \citep{DobricOjeda2009},  at $H \in [0.4,1]$ the DO process  $V_H$ approximates $Z_H$ with a relative $L^2$ error at most at 12\%, At lower $H$ the discrepancy is bigger and can reach 80-100\% at small $H$. Therefore, in \citep{ADOL_Risk}
an adjusted DO (ADO) process is proposed which is defined as
\begin{equation} \label{ADO}
\calVH(t) = \psi_H(t) M_H(t)  + \iu d_H t^H = V_H(t) + \iu d_H t^H,
\end{equation}
\noindent where $\iu$ is an imaginary unit. The ADO process inherits a semi-martingale property from $V_H(t)$ and provides the minimum $\EE[\calY^2(t)]  = \EE [(Z_H(t) - \calVH(t))^2] = 0$. However, this requires an extension of the traditional measure theory into the complex domain, see, e.g., \citep{CarrWu2004}.

As from the definition, $V_H(t) = \calVH(t)  - \iu d_H t^H$, \eqref{DifRepr} can be transformed to
\begin{align} \label{DifRepr1}
d\calVH(t) &= \left[ \iu H d_H t^{H-1} + \dfrac{2H-1}{t} \calVH(t) \right] dt + B_H t^{H-1/2} d W_t,
\end{align}
\noindent with the same BM as in \eqref{DifRepr}. In other words, the ADO process can also be represented as an \Ito diffusion. If $H < 1/2$ it exhibits mean-reversion. However, the ADO process is not a martingale anymore under $\calF_t^H$ due to the adjustment made. However, as we use this process for modeling the instantaneous variance, it should not be a martingale.

\paragraph{The mean-reversion term.} Another modification is about the mean-reversion term of the instantaneous variance process. As compared with the standard Heston model where this term is linear in $v_t$ here we use the representation $\kappa (\theta - \sqrt{v_t})$. Again, this is done for the tractability reason,

\paragraph{The model.} For the easiness of  notation let us use symbols $\calV_t$ instead of $\calVH(t)$, and $\nu(t) = B_H t^{H-1/2}$. Then, assuming real-world dynamics (i.e., under measure $\mathbb{P}$), the ADO-Heston model is defined by the following  system of stochastic differential equations (SDE)
\begin{align} \label{ADOheston3}
dS_t &= S_t\mu dt + S_t \sqrt{v_t} dW^{(1)}_t \\
dv_t &=  [\kappa (\theta(t) - \sqrt{v_t}) + \xi D_v \sqrt{v_t}] dt + \xi \nu(t) \sqrt{v_t} dW^{(2)}_t,  \nonumber \\
d\calV_t &= D_v dt + \nu(t) dW^{(2)}_t, \qquad D_v = \left[ \iu H d_H t^{H-1} + \dfrac{2H-1}{t} \calV_t \right] \ST{t > \epsilon}, \nonumber \\
\langle dW^{(1)}_t, dW^{(2)}_t \rangle &= \rho dt, \qquad S_{t=0} = S, \ v_{t=0} = v, \ \calV_{t=0} = \calV, \quad (t, S_t, v_t) \in [0,\infty), \ \calV \in (-\infty, \infty), \nonumber
\end{align}
\noindent where $W^{(1)}$ and $W^{(2)}$ are two correlated Brownian motions with the constant correlation coefficient $\rho$, $\kappa$ is the rate of mean-reversion, $\xi$ is the volatility of volatility $\sigma_t$ (vol-of-vol), $\theta(t)$ is the time-dependent mean-reversion level (the long-term run), $r$ is the interest rate and $q$ is the continuous dividend and $\mu$ is the drift. All parameters of the model are assumed to be time-independent, despite this assumption could be relaxed,
see, e.g., \citep{Benhamou2010, Rouah2013, CarrItkinMuraveyHeston} and references therein. The process for $\calV_t$ is the OU process with time-dependent coefficients. As by definition in \eqref{ADOheston3} the drift $D_v$ vanishes at $t=0$, the mean-reversion speed of $v_t$ at $t=0$ becomes $-k$, i.e., is well-defined $\forall H \in [0,1]$.

Obviously, the instantaneous variance should not be negative, i.e., $v_t \ge 0$. It has to be checked that the process $v_t$ as it is defined in \eqref{ADOheston3} cannot go negative. Otherwise, some boundary condition should be set at the boundary $v_t = 0$. In more detail this is analysed in Section~\ref{solEq24}.

The ADO-Heston model  is a two-factor model (actually, three stochastic variables $S_t, v_t, \calV_t$ are introduced, but two of them: $v_t$ and $\calV_t$ are fully correlated). As mentioned in \citep{ADOL_Risk}, the model in \eqref{ADOheston3} is a SVM where the speed of mean-reversion of $v_t$ is stochastic, but fully correlated with $v_t$. In the literature there have been already some attempts to consider an extension of the Heston model by assuming the mean-reversion level $\theta$ to be stochastic, see \citep{Gatheral2008,Bi2016}. In particular, in \citep{Gatheral2008} it is shown that such a model is able to replicate a term structure of VIX options. However, to the best of our knowledge, stochastic mean-reversion speed has not been considered yet.

The ADO-Heston process for $v_t$ in a certain sense is similar to that introduced in \citep{StochMR2016} who considered a generalized OU process by letting a mean-reversion speed to be stochastic and, in particular, a Brownian stationary process. As our process $\calV_t$ is also a time-dependent OU process, it may attain negative values, so the mean-reversion rate could become negative. However, in \citep{StochMR2016}, the authors are able to show the stationarity of the mean, the variance, and the covariance of the process (the process $\sigma_t$ in our notation) when the average speed of mean-reversion is sufficiently larger than its variance. Explicit conditions for these results to hold are also derived in that paper.

\section{The ADO-Heston PDE for the option price} \label{ADOHPDE}

To price options written on the underlying stock price $S_t$ which follows the ADO-Heston model, a standard approach can be utilized, \citep{Gatheral2006,Rouah2013}. Consider a portfolio consisting of one option $V = V (S, v, \calV, t)$, $\Delta$ units of the stock $S$, and $\phi$ units of another option $U = U(S, v, \calV, t)$ that is used to hedge the volatility. The dollar value of this portfolio is
\begin{equation} \label{port}
\Pi = V + \Delta S + \phi U.
\end{equation}
The change in the portfolio value $d \Pi$ could be found by applying \Ito  lemma to $dV$ and $dU$, and assuming that the continuous dividends are re-invested back to the portfolio
\begin{align} \label{port2}
d\Pi &= d V + \Delta d S + \phi d U + \Delta q S dt, \\
&= \Bigg\{ \fp{V}{t} + \frac{1}{2} v S^2 \sop{V}{S} + \frac{1}{2} \xi^2 v \nu^2(t) \sop{V}{v} + \frac{1}{2} \nu^2(t) \sop{V}{\calV} \nonumber \\
&\hspace{0.55in} +  \rho S \xi v  \nu(t) \cp{V}{S}{v} + \rho S \sqrt{v} \nu(t) \cp{V}{S}{\calV} + \xi \sqrt{v} \nu^2(t) \cp{V}{\calV}{v} \Bigg\} dt \nonumber \\
&+ \phi \Bigg\{ \fp{U}{t} + \frac{1}{2} v S^2 \sop{U}{S} + \frac{1}{2} \xi^2 v \nu^2(t) \sop{U}{v} + \frac{1}{2} \nu^2(t) \sop{U}{\calV} \nonumber \\
&\hspace{0.55in} + \rho S \xi v  \nu(t) \cp{U}{S}{v} + \rho S \sqrt{v} \nu(t) \cp{U}{S}{\calV} + \xi \sqrt{v} \nu^2(t) \cp{U}{\calV}{v} \Bigg\} dt \nonumber \\
&+ \Bigg\{ \fp{V}{S} + \phi \fp{U}{S} + \Delta \Bigg\} dS + \Bigg\{ \fp{V}{v} + \phi \fp{U}{v} \Bigg\} dv +
\Bigg\{ \fp{V}{\calV} + \phi \fp{U}{\calV} \Bigg\} d \calV + \Delta q S dt. \nonumber
\end{align}
Based on \eqref{ADOheston3}, the last three terms in \eqref{port2} in the explicit form could be re-written as
\begin{align} \label{port1}
\Bigg\{ \fp{V}{S} &+ \phi \fp{U}{S} + \Delta \Bigg\} dS + \Bigg\{ \fp{V}{v} + \phi \fp{U}{v} \Bigg\} dv +
\Bigg\{ \fp{V}{\calV} + \phi \fp{U}{\calV} \Bigg\} d \calV \\
&= \Bigg\{ \fp{V}{S} + \phi \fp{U}{S} + \Delta \Bigg\}\left[S \mu dt + S \sqrt{v} d W^Q_{1,t}\right] \nonumber \\
&+ \Bigg\{ \fp{V}{v} + \phi \fp{U}{v} \Bigg\}\left[\kappa (\theta(t) - \sqrt{v}) + \xi \bar{D}_v \sqrt{v} \right] dt +
\Bigg\{ \fp{V}{\calV} + \phi \fp{U}{\calV} \Bigg\} \bar{D}_v dt \nonumber \\
&+   \nu(t) d W^Q_{2,t} \Bigg\{ \left[\fp{V}{\calV} + \phi \fp{U}{\calV} \right] + \xi \sqrt{v} \left[ \fp{V}{v} + \phi \fp{U}{v}\right] \Bigg\}. \nonumber
\end{align}

To make this portfolio riskless, the risky terms proportional to increments of the Brownian Motions must vanish. This implies that the hedge parameters are
\begin{align} \label{pars}
\Delta &= -\fp{V}{S} - \phi \fp{U}{S}, \\
\phi &= -\left[\xi \sqrt{v} \fp{V}{v} + \fp{V}{\calV} \right]\left[ \xi \sqrt{v} \fp{U}{v} +  \fp{U}{\calV} \right]^{-1}. \nonumber
\end{align}
Also, a relative change of the risk free portfolio is the interest earned with the risk free interest rate, i.e.
\begin{equation} \label{rn}
d \Pi = r \Pi dt.
\end{equation}
With allowance for \eqref{pars}, \eqref{port2} could be represented in the form $d\Pi = (A + \phi B) dt$. Therefore, \eqref{rn} can be transformed to
\begin{equation} \label{rn1}
A + \phi B  = r (V + \Delta S + \phi U).
\end{equation}
Using the definition of $\phi$ in \eqref{pars}, this could be re-written as
\begin{equation} \label{rn3}
\frac{A - r V + (r-q) S \fp{V}{S} }{\xi \sqrt{v} \fp{V}{v} + \fp{V}{\calV} }  = \frac{B - r U + (r-q) S \fp{U}{S}}{\xi \sqrt{v} \fp{U}{v} +  \fp{U}{\calV} }.
\end{equation}
The left-hand side of this equation is a function of $V$ only, and the right-hand side is a function of $U$ only. This could be only if both sides are just some function $f(t,S,v,\calV)$ of the independent variables. Accordingly, using the explicit expression for $A$, from \eqref{rn3} we obtain the ADO-Heston PDE (partial differential equation)
\begin{align} \label{pde}
0 &= \fp{V}{t} + \frac{1}{2} v S^2 \sop{V}{S} + \frac{1}{2} \xi^2 v  \nu^2(t) \sop{V}{v} + \frac{1}{2} \nu^2(t)  \sop{V}{\calV} \\
&+  \rho S \xi v  \nu(t) \cp{V}{S}{v} + \rho S \sqrt{v}\nu(t) \cp{V}{S}{\calV} + \xi \sqrt{v} \nu^2(t) \cp{V}{\calV}{v}   \nonumber \\
&+ (r-q) S \fp{V}{S}  + (\bar{D}_v - f) \fp{V}{\calV} + \left[\kappa (\theta(t) - \sqrt{v}) + \xi (\bar{D}_v-f) \sqrt{v} \right]
\fp{V}{v} - r V. \nonumber
\end{align}
To proceed, we need to choose an explicit form of $f(t,S,v,\calV,t)$\footnote{Since two drivers of our model are not tradable, the market price of risk naturally appears even under risk-neutral measure, see, e.g., \citep{Mandel2015}.}. We consider two options. The first one relies on a tractability argument and suggests choosing $f = \bar{D}_v + \lambda$, where, similar to \citep{Heston:93}, $\lambda$ is the market price of volatility risk and is constant. However, with this choice the risk-neutral drift of $v_t$ becomes $[\kappa (\theta -\sqrt{v}) + \xi \lambda]dt$, i.e., the stochastic variance $v_t$ doesn't depend on $\calV_t$. In such a model only the vol-of-vol term is a function of $t$ and the Hurst exponent $H$, so this is a stochastic volatility model with the time-dependent vol-of-vol. This makes this model not rich enough for our purposes, despite it is tractable. Therefore, in what follows we ignore this choice. For the reference, pricing options using the time-dependent log-normal model can be done similar to \citep{Benhamou2010} where option prices in the time-dependent Heston model were found by using an asymptotic expansion of the PDE in a small vol-of-vol parameter.

Another construction, introduced in this paper is the choice $f = \bar{D}_v + g(t, v, \calV, \lambda)$.  With this definition \eqref{pde} takes the form
\begin{align} \label{pde1}
0 &=  \fp{V}{t} + \frac{1}{2} v S^2 \sop{V}{S} + \frac{1}{2} \xi^2 v^2  \nu^2(t) \sop{V}{v} + \frac{1}{2} \nu^2(t) \sop{V}{\calV}  +  \rho S \xi v  \nu(t) \cp{V}{S}{v} \\
&+ \rho S \sqrt{v}\nu(t) \cp{V}{S}{\calV} + \xi \sqrt{v} \nu^2(t) \cp{V}{\calV}{v} + (r-q) S \fp{V}{S} \nonumber \\
&- g(t, v, \calV, \lambda)\fp{V}{\calV} + [\kappa(\theta(t) - \sqrt{v}) - \xi g(t, v, \calV, \lambda) \sqrt{v}] \fp{V}{v} - r V. \nonumber
\end{align}

As by Girsanov's theorem, \citep{karatzas1991brownian}
\begin{align}
d W^{(1)}_t &= d W_{1,t}^{Q} - \gamma_1(t) dt, \\
d W^{(2)}_t &= d W_{2,t}^{Q} - \gamma_2(t) dt, \nonumber
\end{align}
\noindent with $W^Q, W_2^Q$ be the corresponding Brownian motions under measure $\mathbb{Q}$, a necessary condition for this measure to exist is
\[ \mu - (r-q) = \sqrt{v_t} \left( \rho \gamma_2(t) + \sqrt{(1-\rho^2} \gamma_1(t)\right),
\]
\noindent which ensures that the discounted stock price is a local martingale under measure $\mathbb{Q}$, see e.g., \citep{WongHeyde2006}. Accordingly, by using the same argument, one can see that the PDE in \eqref{pde1} corresponds to the following model under the risk-neutral measure $\mathbb{Q}$
\begin{align} \label{ADOhestonQ}
dF_t &=  F_t \sqrt{v_t} dW^Q_{1,t} \\
d v_t &= [\kappa(\theta(t) - \sqrt{v_t}) - \xi g(t, v_t, \calV_t, \lambda) \sqrt{v_t} ] dt +  \xi \nu(t) \sqrt{v_t} dW^{Q}_{2,t},  \nonumber \\
d\calV_t &= - g(t, v, \calV, \lambda) dt + \nu(t) dW^Q_{2,t} , & \nonumber \\
\langle dW^Q_{1,t}, dW^Q_{2,t} \rangle &= \rho dt, \qquad F_{t=0} = F, \quad v_{t=0} = v, \quad \calV_{t=0} = \calV, \nonumber
\end{align}
\noindent where the forward price $F_t$ is introduced instead of the spot $S_t$.  When this model is used for option pricing, and with parameters obtained by calibration of the model to market options prices, one is already in the risk-neutral setting. Then, as explained in \citep{Gatheral2006}, that allows setting the market price of volatility risk $\lambda$ equal to zero. So, in what follows we set $\lambda = 0$.

\section{The CF of the log-price and the ATM implied skew} \label{CFT}

One of the main reasons for popularity of the Heston model is that the characteristic function (CF) of $\log F_T$ in this model is known in closed form. Then any FFT based method, \citep{CarrMadan:99a, Lewis:2000, Lipton2001, FangOosterlee2008}, can be used to price European, and even American, \citep{LordAmerican2007}, options written on the underlying stock $S_t$ or forward $F_t$.

Recall, that in this paper we look at the behavior of the ATM implied skew in the ADO-Heston and similar models. A straightforward but rather naive approach to find the ATM implied skew would be: first, solving the PDE in \eqref{pde1}, then equating the result to the corresponding Black-Scholes (BS) price which is a function of the implied volatility $I$ and solving this new algebraic equation, and finally computing the first derivative of $I$ on the normalized strike $k =\log(F/K)$, and set $k = 0$ to get the ATM value. Obviously, this approach is not much tractable and therefore, a small trick can significantly simplify finding the ATM implied skew. The main idea consists in the fact that the CF of the ADO-Heston model can be determined in closed form under some assumptions, \citep{ADOL_Risk}. And for the BS model it is also known in closed form. Therefore, one can equate the BS and ADO-Heston prices both expressed via their CFs and then solve this equation with respect to the unknown implied volatility $I$. Once the authors came to this idea, they immediately discovered that it had been already elaborated in \citep{Gatheral2006}. The result for an arbitrary model (with zero instantaneous interest rate and dividends) reads
\begin{equation} \label{skewATM}
\left.\frac{\partial I}{\partial k}\right|_{k=0}=-e^{\frac{I^2 T}{8}} \sqrt{\frac{2}{\pi}} \frac{1}{\sqrt{T}} \int_{0}^{\infty} d u \frac{u \operatorname{Im}\left[\phi(u-\iu / 2)\right]}{u^{2} + 1/4},
\end{equation}
\noindent where $\phi(u)$ is the part of the CF of the model which doesn't depend on $F_T$. Thus, to determine the ATM implied skew we need to know the CF of the ADO-Heston model. In \citep{ADOL_Risk} it has been obtained asymptotically assuming small vol-of-vol $\xi$. Here we improve this result and demonstrate that a closed-form expression for the CF can be obtained for arbitrary $\xi$.

Let us pick the following representation of the characteristic function
\begin{equation}
\EE[e^{i u \log F_T} | S, v, \calV]  = e^{i u \log F} \psi(u; x,v,\calV, t),
\end{equation}
\noindent where $\psi(u; x,v,\calV, t) = \EE[e^{i u \log x}]$ and $x = \log F_T/F$.  By the same argument, \citep{ContVolchkova2005}, the CF of the log forward price $F_T$ - solves a PDE similar to \eqref{pde1} but with no discounting term $r V$
\begin{align} \label{pde11}
0 &=  \fp{\psi}{t} +   \frac{1}{2} v \sop{\psi}{x} + \frac{1}{2} \xi^2 v^2 \nu^2(t) \sop{\psi}{v} + \frac{1}{2} \nu^2(t) \sop{\psi}{\calV}  +  \rho \xi v^{3/2} \nu(t) \cp{\psi}{x}{v} + \rho \sqrt{v} \nu(t) \cp{\psi}{x}{\calV} \\
&+ \xi v \nu^2(t) \cp{\psi}{\calV}{v} -  \frac{1}{2}v \fp{\psi}{x}  - g(t, v, \calV, 0) \fp{\psi}{\calV} + [\kappa (\theta  - v) - \xi g(t, v, \calV, 0) v] \fp{\psi}{v}, \nonumber
\end{align}
\noindent with the initial condition $\psi(u; x, v,\calV, T) = 1$. We will search the solution of this PDE in the form
\begin{equation} \label{subst0}
\psi(u; x, v,\calV,t) = e^{\iu u x} \phi(u; t, v, \calV),
\end{equation}
\noindent where $\phi(u; t, \sigma, \calV)$ is a new dependent variable. Substituting \eqref{subst0} into \eqref{pde11} yields
\begin{align} \label{pdePsi1}
0 &=  \fp{\phi}{t} + \frac{1}{2} \xi^2 \nu^2(t) v \sop{\phi}{v} + \frac{1}{2} \nu^2(t) \sop{\phi}{\calV} + \xi \nu^2(t) \sqrt{v} \cp{\phi}{\calV}{v} + \Big[\kappa (\theta(t) - \sqrt{v})  \\
&- \xi g(t, v, \calV, 0) \sqrt{v} + \iu u \rho \xi \nu(t) v \Big] \fp{\phi}{v} + \left[\iu u \rho \nu(t) \sqrt{v} - g(t, v, \calV, 0) \right] \fp{\phi}{\calV} -\frac{1}{2} u (\iu + u) v \phi, \nonumber
 \end{align}
Again, this equation should be solved  subject to the initial condition $\phi(u; T, v, \calV) = 1$.

\subsection{Solution of \eqref{pdePsi1}} \label{solEq24}

To solve \eqref{pdePsi1} we make another simplification and set
\begin{equation} \label{theta}
\theta(t) = \frac{\xi^2 }{4 k}\nu(t)^2.
\end{equation}
The reason for this assumption is again tractability.  By doing so, we intend to show (analytically) that under this assumption an asymptotic behavior of $I$ at $T \to 0$ is same to what  is predicted by rough volatility models. If this is true, then it would be naturally to relax this assumption and again consider an asymptotic behavior of $I$ at $T \to 0$, but now using numerical methods with a hope that this doesn't alter the conclusion.

On the other hand, this assumption makes the mean-reversion level $\theta(t)$ to be a function of the Hurst exponent $H$. We will discuss this in more detail in Section~\ref{disc}.

To proceed, in \eqref{pdePsi1} we make a change of independent variables $\calV \mapsto h = \xi \calV - 2\sqrt{v} + \kappa(T-t)$, so $\phi(u; t,v,\calV) \mapsto z(u; t,v,h)$. Also, we make a particular choice of the function $g(t,v,\calV,0)$ (which now becomes $g(t,v,h,0)$)
\begin{equation} \label{gDef}
g(t,v,\calV(t,v,h,0) = - \frac{\kappa}{\xi} + \dfrac{m(t,h,H) + \xi \nu^2(t)/4}{\sqrt{v}},
\end{equation}
\noindent where $m(t,h,H)$ is some function of the time $t$, new variable $h$ and the Hurst exponent $H$. After that \eqref{pdePsi1} takes the form
\begin{align} \label{pdeP1}
0 &=  \fp{z}{t} + \frac{1}{2} \xi^2 \nu^2(t) v \sop{z}{v} + \xi \left[\iu u \rho \nu(t) v - m(t,h,H)\right]\fp{z}{v} - \frac{1}{2} u (\iu + u) v z,
 \end{align}
 \noindent again to be solved subject to the terminal condition $z(u; T,v,h) = 1$. It can be seen, that despite $z(u; t,v,h)$ is a function of three variables, \eqref{pdeP1} is a two-dimensional PDE with respect to variables $(t,v)$ (so $h$ is a dummy variable). Moreover, it is affine in $v$ since all its coefficients are linear functions of $v$.

Using \eqref{pdeP1} it can be verified that, according to Feller's classification \citep{Feller1952}, the boundary $v=0$ is the entrance point, and thus inaccessible.

Since in \eqref{skewATM} we need function $\phi(u-\iu/2)$, in \eqref{pdeP1} we also switch from $u$ to $u-\iu/2$ that yields
\begin{align} \label{pdeP2}
0 &=  \fp{z}{t} + \frac{1}{2} \xi^2 \nu^2(t) v \sop{z}{v} + \xi \left[\frac{1}{2} (1 + 2 \iu u) \rho \nu(t) v - m(t)\right] \fp{z}{v} - \frac{1}{2} v (u^2 + 1/4) z.
 \end{align}

It is worth noting that in case $\rho = 0$  the solution of \eqref{pdeP2} is a real function (with no imaginary part).\footnote{In case the model is written in terms of the spot price $S_t$, we also need the additional condition $r = q$.} This means that the ATM implied skew in this case is zero (smile is symmetric), indeed, a well-known fact for an  uncorrelated SV model.

The \eqref{pdeP2} is affine in $v$, hence we can represent its solution in the form
\begin{equation} \label{zDef}
z(u; t,v,h) = e^{A(u; t,h) + v B(u; t,h)},
\end{equation}
\noindent where functions $A(u; t,h), B(u; t,h)$ solve the following system of ordinary differential equations (ODE)
\begin{align} \label{ric}
0 &= B'_t(u; t,h) + \frac{1}{2} \xi \rho  (1+2 \iu u) \nu(t) B(u;t,h)  + \frac{1}{2} \xi^2 \nu^2(t) B^2(u; t,h) - \frac{1}{2} (u^2 + 1/4), \\
A'_t(u;t,h) &=  \xi m(t) B(u; t,h),
\qquad B(u;T,h) = A(u;T,h) = 0. \nonumber
\end{align}

The first equation is of the Riccati type (as usual for the Heston model) and can be solved independently. Then solving the first equation is straightforward. Accordingly, the term $\operatorname{Im}\left[\phi(u-i / 2)\right]$ in the integral in \eqref{skewATM} reads
\begin{align} \label{ImInteg}
\operatorname{Im}\left[\phi(u-i / 2)\right] &=
e^{A_R(u;t,h)  + v B_R(u;t,h)]} \sin \left[A_I(u;t,h)  + v B_I(u;t,h)\right], \\
A_R(u;t,h)  + v B_R(u;t,h) &= v B_R(u;t,h) +  \xi \int_t^T m(s) B_R(u;s,h) ds, \nonumber \\
A_I(u;t,h)  + v B_I(u;t,h) &= v B_I(u;t,h) + \xi \int_t^T m(s) B_I(u;s,h) ds, \nonumber \\
B_R(u;t,h) &\equiv \operatorname{Re}[B(u;t,h)], \qquad B_I(u;t,h) \equiv \operatorname{Im}[B(u;t,h)]. \nonumber
\end{align}

\subsection{Solution of the Riccati equation} \label{ricSect}

By a standard change of variables
\begin{align} \label{transRic}
s(t) &= q_2(t) q_0, \qquad p(t) = q_1(t) + \frac{q'_2(t)}{q_2(t)}, \qquad B(u; t,h) = - \frac{w'(t)}{q_2(t) w(t)}, \\
q_2(t) &= 	-\frac{1}{2} \xi^2 \nu^2(t), \quad q_1(t) = -\frac{1}{2} \xi \rho  (1+2 \iu u) \nu(t), \qquad q_0 = \frac{1}{2} (u^2 + 1/4), \nonumber
\end{align}
\noindent the first line in \eqref{ric} can be transformed to a linear ODE
\begin{equation} \label{linODE}
	w''(t) - p(t) w'(t) + s(t) w(t) = 0.
\end{equation}
Then the solution of \eqref{linODE} can be obtained in closed form in terms of the Kummer functions, \citep{as64}. However, first, it is bulky, and second, we are interested in the behavior of this solution at $T \to t$. Therefore, instead for solving \eqref{ric} we will use asymptotic expansions. By the terminal condition in \eqref{ric} we have $B(u;T,h) = 0$, and substituting it into \eqref{ric}: $B'(u;T,h) = \frac{1}{2} (u^2 + 1/4)$. Differentiating \eqref{ric} by $t$ and setting $t \to T$ we also obtain $B''(u;T,h) = -\frac{1}{4} \xi  \rho  (1+2 \iu u) \left(u^2+\frac{1}{4}\right) \nu(T)$. Overall, this gives rise to the following representation
\begin{equation} \label{series}
B(u;t,h) = \frac{1}{2} \left(u^2+\frac{1}{4}\right) (t-T) \left[ 1 - \frac{1}{4} \xi  \rho  (1+2 \iu u) \nu(T)^2 (t-T)\right] + O((t-T)^3).
\end{equation}
This representation solves the Riccati equation in \eqref{ric} with the accuracy $O((t-T)^2)$. It can be seen that the leading term of the real part of $B(u;t,h)$ is $\Ree(B(u;t,h)) \propto O(t-T)$ while $\Imm(B(u;t,h)) \propto O((t-T)^2)$.

Note, that since $t \le T$ we have $\operatorname{Re} B(u;t,h) \le 0$ if $\rho > 0$. Therefore, the solution in \eqref{zDef} behaves well at $v \to \infty$. For $\rho < 0$ we need to take into account the next term in the series, i.e., the term which is $O((t-T)^3)$.

Accordingly, the term $\operatorname{Im}\left[\phi(u-i / 2)\right]$ in the integral in \eqref{skewATM} reads
\begin{align} \label{ImInteg2}
\Imm \left[\phi(u-i / 2)\right] &= e^{A_R(u;t,h)  + v B_R(u;t,h)]} \sin \left\{A_I(u;t,h)  + v B_I(u;t,h)]\right\}, \\
B_R(u;t,h) &= \frac{1}{2} \left(u^2+\frac{1}{4}\right) (t-T) \left[ 1 - \frac{1}{4} \xi  \rho \nu(T)^2 (t-T)\right], \nonumber \\
B_I(u;t,h) &= -\frac{1}{4} u \left(u^2+\frac{1}{4}\right) \xi  \rho  \nu(T)^2 (t-T)^2, \nonumber \\
A_R(u;t,h) &= - \xi \int_t^T m(s) B_R(u;s,h) ds, \qquad A_I(u;t,h) = -\xi \int_t^T m(s) B_I(u;s,h) ds. \nonumber
\end{align}

\subsection{Choice of $m(t,h,H)$} \label{example}

For modeling the observable behavior of the implied vanilla or forward skew the choice of function $m(t,h,H)$ is important.  In what follows we set it as (compare with the drift in \eqref{DifRepr1})
\begin{equation} \label{mDef}
m(t,h,H) = -\frac{\zeta(h)}{\xi t^{1-H}}\ST{t > \epsilon},
\end{equation}
\noindent where $\zeta(h)$ is some arbitrary function of $h$. Here the principal point is the time dependence of $m(t,h,H)$ as a power function of $t$, i.e. $m(t,h,H) \propto t^{H-1}$. Otherwise the short time behavior of the implied skew at $T \to t$: $\left. (\partial I/\partial k)\right|_{k=0} \propto T^{H-1/2}$  either cannot be replicated at all, or the approximation behaves worse as compared with what we propose here.

Using this definition of $m(t,h,H)$  and setting $t \to 0$ we obtain up to the leading terms in $T \ll 1$
\begin{align} \label{ImInteg3}
A_R(u;t,h) &= -\frac{1}{12} \zeta(h)  \left(u^2+\frac{1}{4}\right) \left[(5 - 2 H)T^{H+1} + \frac{1}{2} B^2_H \rho \xi u T^{3H+1} \right],  \\
A_I(u;t,h) &= -\frac{1}{12} \zeta(h)B^2_H \xi \rho  u \left(u^2+\frac{1}{4}\right) T^{3 H+1}. \nonumber
\end{align}
Since from \eqref{ImInteg2} at $ t=0$ and $T \to t$ we have $B_R(u;t,h) \propto O(T), \ B_I(u;t,h) \propto  O(T^2)$, from \eqref{ImInteg2} up to the leading terms in small $T$ we have
\begin{equation} \label{phiSol}
\Imm \left[\phi(u-i / 2)\right] = e^{A_R(u;t,h) + I^2 B_R(u;t,h)} \sin [A_I(u;t,h)].
\end{equation}

\subsection{Computation of the integral in \eqref{skewATM}}

Using the definition in \eqref{skewATM} and the representation of $\Imm \left[\phi(u-i / 2)\right]$ in \eqref{phiSol}, we can now write the final asymptotic expression for $\left. (\partial I/\partial k)\right|_{k=0}$ at  $T \to t$
\begin{equation} \label{skewATMFin}
{\cal S}_{anal} = \left.\frac{\partial I}{\partial k}\right|_{k=0}=-e^{\frac{I^2 T}{8}} \sqrt{\frac{2}{\pi}} \frac{1}{\sqrt{T}} \int_{0}^{\infty} d u \frac{u\, e^{A_R(u;t,h) + I^2 B_R(u;t,h)} \sin [A_I(u;t,h)]}{u^{2} + 1/4}.
\end{equation}

\begin{figure}[!htb]
\captionsetup{format=plain}
\centering
\includegraphics[width=0.7\textwidth]{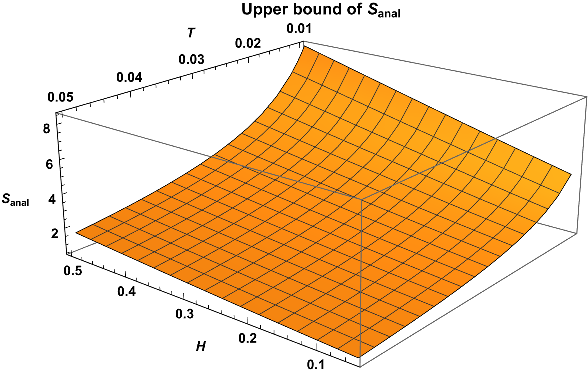}
\caption{The upper bound of the implied skew ${\cal S}_{anal}$ computed by using \eqref{skewATMFin} as a function of $T, H$. }
\label{Fig0}
\end{figure}

The upper bound of ${\cal S}_{anal}$ can be found analytically by setting $\sin [A_I(u;t,h)] = -1$, because then the integral in the RHS of \eqref{skewATMFin} can be found in closed form
\begin{align} \label{skewATMFinB}
{\cal S}_{anal} &= e^{\frac{I^2 T}{8}} \frac{1}{\sqrt{2 \pi T}} \Gamma \left(0,p(T,H)\right), \\
p(T,H) &= \frac{1}{12} \zeta(h) \left[(5 - 2 H)T^{H+1} + \frac{1}{2} B^2_H \rho \xi T^{3H+1} \right] + \frac{1}{2} I^2 T, \nonumber
\end{align}
\noindent where $\Gamma(a, z)$ is the incomplete gamma function, \citep{as64}. The plot of the RHS of \eqref{skewATMFinB} with the model parameters given in Table~\ref{param} is depicted in Fig.~\ref{Fig0}
It can be seen that, indeed ${\cal S}_{anal}$ grows with $T \to 0$

\begin{figure}[!htb]
\captionsetup{format=plain}
\centering
\subfloat[]{\includegraphics[width=0.49\textwidth]{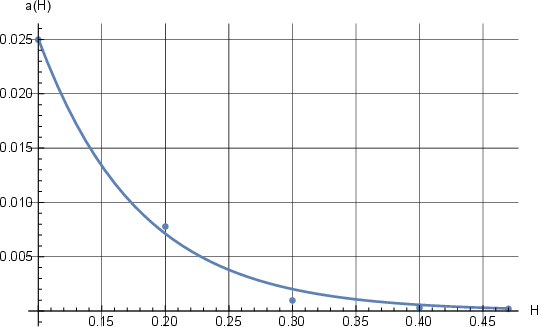}}
\subfloat[]{\includegraphics[width=0.49\textwidth]{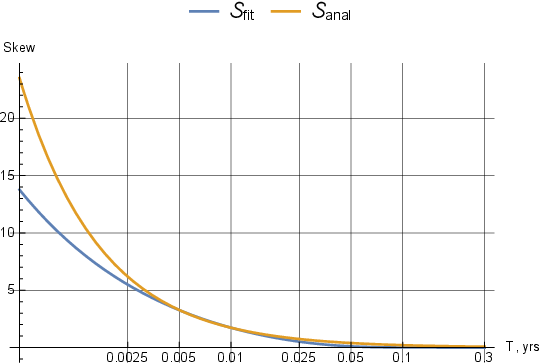}} \\
\subfloat[]{\includegraphics[width=0.49\textwidth]{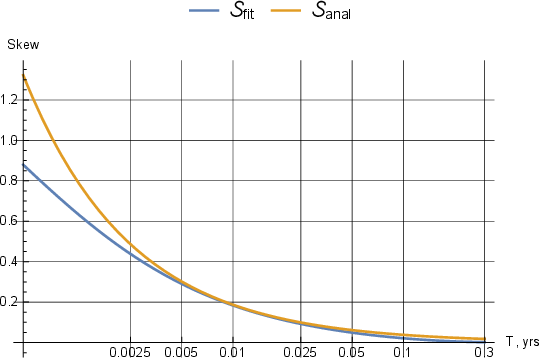}}
\subfloat[]{\includegraphics[width=0.49\textwidth]{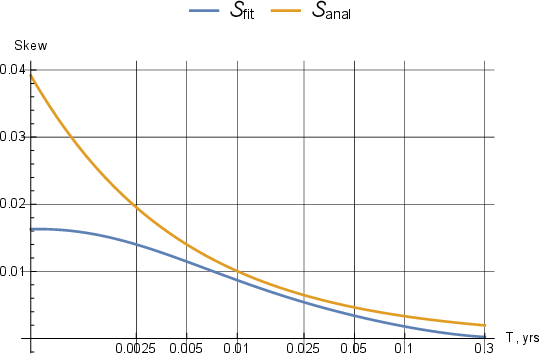}} \\
\subfloat[]{\includegraphics[width=0.49\textwidth]{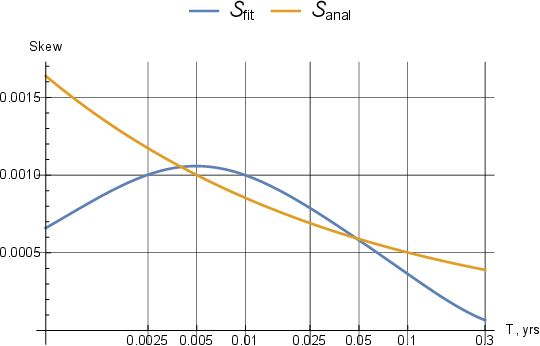}}
\subfloat[]{\includegraphics[width=0.49\textwidth]{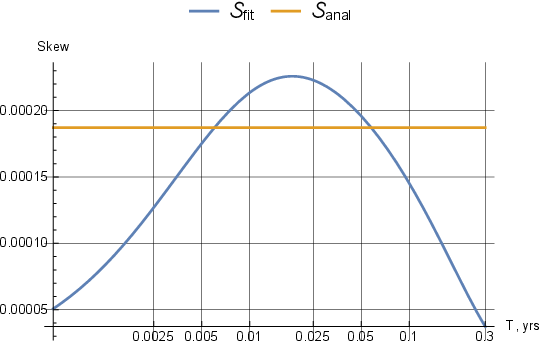}} \\
\caption{The implied skew ${\cal S}_{anal}$ computed by using \eqref{skewATMFin} vs the function ${\cal S}_{fit}$ in a test with parameters given in Table~\ref{param}. Here: a)a regressed dependence a(H) vs the data points in Table~\ref{aofH}; b) - f) comparison of two skews for $H=0.1,0.2,0.3,0.4,0.5$ respectively. }
\label{Fig1}
\end{figure}

For a more accurate estimation of ${\cal S}_{anal}$ let us again set some values of the model parameters, compute this integral numerically as a function of $T$ and regress it to the function $a(H) T^{b(H)}$, where $a(H), b(H)$ are the weights to be determined. The values of the model parameters used in our experiment are given in Table~\ref{param}
\begin{table}[!htb]
\begin{center}
\begin{tabular}{|c|c|c|c|c|c|c|c|c|}
\hline
$I$ & $\rho$ & $\xi$ & $\zeta(h)$ \\
\hline
0.5 &  0.7 & 0.01 & 100 \\
\hline
\end{tabular}
\caption{Parameters of the test.}
\label{param}
\end{center}
\end{table}
Computing the integral in \eqref{skewATMFin} for $T \in [0.001,0.3]$ and $H \in [0.1,0.5]$ and regressing thus obtained data gives rise to an approximate dependence ${\cal S}_{fit} \propto a(H) T^{2.3(H-1/2)}$. Here the values of $a(H)$ are given in Table~\ref{aofH} and could be regressed as, e.g., $a(H) = e^{-12.5927 H-2.42651}$.
\begin{table}[!htb]
\begin{center}
\begin{tabular}{|c|c|c|c|c|c|c|c|c|}
\hline
$H$ & 0.1 & 0.2 & 0.3 & 0.4 & 0.47 & 0.5  \\
\hline
a(H) &  0.02498 & 0.00778 & 0.00098 & 0.00030 & 0.00020 & 0.00019 \\
\hline
\end{tabular}
\caption{The discrete function $a(H)$ found by regressing ${\cal S}_{anal}$ to ${\cal S}_{fit}$. }
\label{aofH}
\end{center}
\end{table}
The results of this test are also presented in Fig.~\ref{Fig1}.

It can be seen that for maturities below 3 months and up to 1 day (which is, perhaps, a good practical lower limit)  the function ${\cal S}_{fit}$ is close to ${\cal S}_{anal}$ for $H[0.1,0.3]$. For $H \in [0.4,0.5]$ the difference is more pronounced, however, the value of the skew is small. Also, ${\cal S}_{anal}$  at very small $T < 0.025$ goes down when $H \in [0.4,0.5]$ while ${\cal S}_{fit}$ goes up. Also, in ${\cal S}_{fit}$  the power of $T$ is not $H-1/2$ as in rough volatility models but is approximately proportional to $H-1/2$.

To recall, ${\cal S}_{anal}$ is just a low limit approximation (up to $O(T^2)$) of the exact solution (which can be obtained by solving the PDE in \eqref{pdePsi1} and then computing the integral in \eqref{skewATM}). Nevertheless, the claim can be made that the ADO-Heston model (which is a pure diffusion model but with a stochastic mean-reversion speed of the variance process, or a Markovian approximation of the rough Heston model) approximately is able to reproduce the known behavior of the vanilla implied skew at small $T$.

\section{Preliminary discussion} \label{disc}

On the way to obtain the final representation of the implied skew ${\cal S}(T)$ we made various simplifications, mostly by a tractability argument, with the belief that the original model with no simplifications could produce similar results (despite, most likely,  they can be obtained only numerically). Here we want to collect all those simplifications together and discuss them to make our ADO-Heston model as much transparent as possible. Using the definitions in \eqref{mDef}, \eqref{gDef}, \eqref{theta} and substituting them into \eqref{ADOhestonQ} we obtain
\begin{align} \label{ADOhestonFin}
dF_t &=  F_t \sqrt{v_t} dW^Q_{1,t} \\
d v_t &= \zeta(h_t) t^{H-1}\ST{t > \epsilon} dt  +  \xi \nu(t) \sqrt{v_t} dW^{Q}_{2,t},
\qquad h_t = \xi \calV_t - 2\sqrt{v_t} + \kappa(T-t), \nonumber \\
d\calV_t &= \frac{\beta(t,v,\calV)}{\xi} dt + \nu(t) dW^Q_{2,t},
\qquad \beta(t, v, \calV) = \kappa - \frac{\xi}{4 \sqrt{v}}\left[ \xi \nu^2(t) + 4 \zeta(h) t^{H-1}\ST{t > \epsilon} \right],  \nonumber \\
\langle dW^Q_{1,t}, dW^Q_{2,t} \rangle &= \rho dt, \qquad F_{t=0} = F, \quad v_{t=0} = v, \quad \calV_{t=0} = \calV, \nonumber
\end{align}
In the limit of $T \to t$ the last term in the definition of $h$ can be neglected. Then, taking, e.g., $\zeta(h) = \alpha  h, \ \alpha - const$, one can expect that such a model preserves (under an appropriate choice of the model parameters) mean-reversion of the instantaneous variance $v_t$. The speed of this mean-reversion is inversely proportional to $t^{1-H}$, so it is high at small $t$. However, there is no singularity in the definition of $\beta(t.v.H)$ since both terms in square brackets are integrable functions of $t$. The mean-reversion level in such a model is stochastic and determined by the stochastic variable $\calV_t$. The SDE for $\calV_t$ also has (potentially) a mean-reverting drift, now in the variable $\calV$. Both the drift and volatility of $v_t$ and the drift and volatility of $\calV_t$ are functions of the Hirst exponent $H$. In turn, the mean-reversion level of $\calV_t$ is also stochastic and inversely proportional to $\sqrt{v}_t$.

\begin{figure}[!htb]
\centering
\fbox{
\subfloat[]{\includegraphics[width=0.49\textwidth]{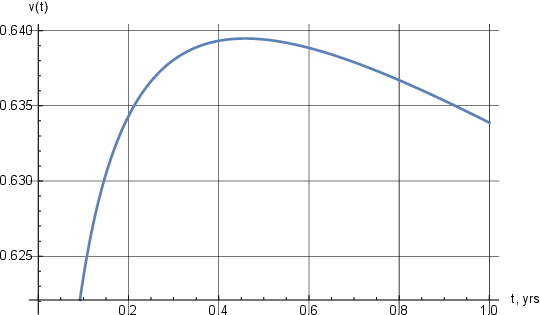}}
\subfloat[]{\includegraphics[width=0.49\textwidth]{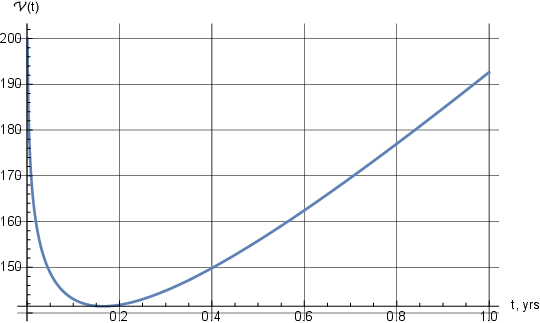}}
}\\
\caption{Plots of functions $v(t)$ - a, and $\calV(t)$ - b, computed by using the deterministic ODEs (the 2nd and 3rd ones) in \eqref{ADOhestonFin} and parameters given in Table~\ref{Tab2}.}
\label{Test2}
\end{figure}

To illustrate this analysis, let us consider a deterministic version of the second and third SDEs in \eqref{ADOhestonFin}, i.e. with drifts but with no stochastic terms. We use the values of the model parameters given in Table~\ref{Tab2} and then solve this system of two ODEs numerically.
\begin{table}[!htb]
\begin{center}
\begin{tabular}{|c|c|c|c|c|c|c|c|c|}
\hline
$v_0$ & $\calV_0$ & $H$ & $\kappa$ & $\xi$ & $T$ & $\alpha$ \\
\hline
0.5 & 200 & 0.1 & 1 & 0.01 & 1 & 0.1\\
\hline
\end{tabular}
\caption{Parameters of the mean-reversion test.}
\label{Tab2}
\end{center}
\end{table}
Thus obtained graphs of the deterministic functions $v(t), \calV(t)$ are presented in Fig.~\ref{Test2}.
Thus, in this case $v(t)$  has a maximum which could be associated with a mean-reversion level. However, after reaching this level the variance doesn't stay at it, but instead drops down with some speed. This occurs because the mean-reversion level is stochastic as this was already explained. A similar behavior, but with a minimum value, can be observed for $\calV(t)$.

Let us shortly summarize our results obtained so far in this paper.
\begin{enumerate}
\item We propose a Markovian approximation of the rough Heston model doing this in the spirit of \citep{ADOL_Risk}. We call it as the ADO-Heston model. Under the risk-neutral measure it is specified by SDEs in \eqref{ADOhestonQ}. There is a minor difference with the Heston model in the drift of the instantaneous variance $v_t$ since we use the mean-reversion term in the form $\kappa(\theta(t) - \sqrt{v_t})$ rather than $\kappa(\theta - v_t)$ in the homogeneous Heston model. We also write our model for the forward price $F_t$.

\item By assuming a special form of the mean reversion level $\theta(t)$ given in \eqref{theta} and using a special form of the market price of risk we managed to find a closed form solution for the characteristic function of the log-price. This solution is expressed via a function $B(u;t,h,H)$ which solves the Riccati equation in \eqref{ric}.

\item Since we are interested in the behavior of the ATM implied skew ${\cal S}$ when the time to maturity $T$ tends to $t$ (so $|t-T| \ll 1$), this Riccati equation can be solved asymptotically. After that a closed form representation of the implied skew is obtained by using \eqref{skewATM}.

\item As a test we then choose a set of the model parameters and show that the behavior of ${\cal S}(T)$ at small $T$ can approximately replicate that one known for rough volatility models, i.e.  ${\cal S}(T) \propto a(H) T^{b(H-1/2)}$. However, for our model $b \ne 1$. In other words, the behavior of our curve is not exactly same as in rough volatility models but seems to be close enough for all practical values of $T$. Thus, the proposed Markovian model is able to replicate some properties of the corresponding rough volatility model. This is important since a Markovian model can be efficiently solved numerically, e.g., for option pricing, by using proven and fast finite-difference or radial basis functions methods.

\item  We also show that the ADO-Heston model could preserve mean-reversion of the instantaneous variance which is also an important property justified by the market. However, in our case this mean-reversion has a more complicated behavior.

\end{enumerate}

Now, a natural question would be about simplifications made to make the model tractable, namely: if one relaxes these simplifications will the model still preserve those nice properties of the corresponding rough model. Most likely, the answer is positive since the only really important simplification is a special form of the mean-reversion level $\theta(t)$. It can be relaxed, and then the implied ATM skew can be computed numerically by solving the PDE in  \eqref{pdePsi1}. Our intuition tells us that relaxing this assumption doesn't significantly change the results, i.e. the behaviour of ${\cal S}(T)$ at small $T$, but we plan to fulfil this program and justify this in our future research.

\section{Forward started options} \label{fso}

A forward started option is a variant of a standard European (vanilla) option where, however, it is purchased and paid at time $t \ge 0$ (now) but becomes active later at time $s > t, \ s \in [0,T)$ with a strike price determined at that time. Hence, this option becomes path-dependent, and is related to exotic rather than vanilla options. The traded options might have two types of payoffs at maturity. For instance, for the forward Call option they are $C(s,T,S_T) = (S_T - K S_s)^+$ or $C(s,T,S_T) = (S_T/S_s - K)^+$, where $K$ is the strike (now dimensionless, and in a sense of the European vanilla options, $K_{vanilla}$ is now proportional to the option price, i.e. $K_{vanilla} = K S_s$). In the following for the sake of certainty we will use the latter definition which, however, doesn't bring any restriction and can be relaxed.

To proceed with the analysis of the implied volatility skew similar to that provided for the vanilla options, we utilize the same approach as in the previous sections. Namely, again we  want to derive an explicit representation of the skew (in spirit of \eqref{skewATM}) where now $\phi(u)$ should be a part of the {\it forward CF} of the model. Thus, to determine the ATM implied skew we need to know the forward CF of the ADO-Heston model and that one for the Black-Scholes model. For doing that we also need to define what is the ATM strike for the forward starting options. A natural choice would be to set $K=1$ because then $K_{vanilla} = S_s$. Therefore, in what follows we will use this definition. Then, the following proposition holds
\begin{proposition} \label{prop1}
The forward ATM implied skew (e.g., the ATM implied skew computed for the forward starting options) is given by the formula similar to \eqref{skewATM} where now instead of the characteristic function $\phi(u)$ one has to use the forward characteristic function $\phi_{s, T}(u)$ defined as, \citep{hong2004}
\begin{equation}
\phi_{s,T}(u) = \mathbb{E}_{\mathbb{Q}}\left[\exp \left(\iu u \cdot \calS_{s, T}\right) \mid S_0, v_0, \calV_0\right] \equiv \int_{-\infty}^{\infty} e^{\iu u \eta} q_{s, T}(\eta) d\eta.
\end{equation}
\noindent Here, $\calS_{s,T} = \log(S_T/S_s)$, $k = \log(K)$, $q_{s,T}(x)$ is the risk-neutral density of the log price $\calS_{s,T}$. The result reads
\begin{equation} \label{skewATMForward}
\left.\frac{\partial I}{\partial k}\right|_{k=0}=-e^{\frac{I^2 (T-s)}{8}} \sqrt{\frac{2}{\pi(T-s)}} \int_{0}^{\infty} d u \frac{u \operatorname{Im}\left[\phi_{s,T}(u-i / 2)\right]}{u^{2} + 1/4},
\end{equation}

\begin{proof}[{\bf Proof}]
We present just a sketch of the proof since \eqref{skewATMForward} can be derived based on the known results. Indeed, first observe that for the Black-Scholes model there exists a closed form representation of the forward starting Call option price, see e.g., \citep{hong2004} among others
\begin{equation} \label{BSforward}
C_{s,T}(K,I) = e^{-r s} C(K, 1, T-s, I),
\end{equation}
\noindent where $C(K, S, \tau, \sigma)$ denotes the Black-Scholes formula, \citep{hull2011}. This equation can be re-written by using another representation given in \citep{hong2004} based on the classic FFT approach of \citep{CarrMadan:99a}, which for an arbitrary model provides the forward starting Call option price in the form
\begin{align} \label{hongCall}
C(s, T, K) &= e^{-r T} \frac{\exp (-\alpha k)}{2\pi} \int_{-\infty}^{\infty} e^{-\iu v k} \frac{\phi_{s, T}(v - (\alpha + 1)\iu)}{(\alpha + \iu v)(\alpha + 1 + \iu v)} dv,
\end{align}
\noindent where $\alpha$ is the dumping factor. Combining \eqref{BSforward} and \eqref{hongCall} we obtain for the Black Scholes model
\begin{align} \label{CallFviaFFT}
C(s, T, K) &= e^{-r T} \frac{\exp (-\alpha k)}{2\pi} \int_{-\infty}^{\infty} e^{-\iu v k} \frac{\phi_{s, T}(v - (\alpha + 1)\iu)}{(\alpha + \iu v)(\alpha + 1 + \iu v)} dv = e^{-r s} C(K, 1, T-s, I) \\
&= e^{-r T} \frac{\exp (-\alpha k)}{2\pi} \int_{-\infty}^{\infty} e^{-\iu v k} \frac{\phi_{T-s}(v - (\alpha + 1)\iu)}{(\alpha + \iu v)(\alpha + 1 + \iu v)} dv, \nonumber \\
\phi_{T}(u) &\equiv \phi(u) = \exp \left[\iu u \left(r-\delta - \frac{1}{2} I^2 \right)T - \frac{1}{2}u^2 I^2 T\right]. \nonumber
\end{align}
From \eqref{CallFviaFFT} we immediately get that for the Black-Scholes model
\begin{align}
\phi_{s, T}(u) = \exp \left[\iu u \left(r-\delta - \frac{1}{2} I^2 \right)(T-s) - \frac{1}{2}u^2 I^2 (T-s)\right].
\end{align}

Now, four important points have to be taken into account
\begin{enumerate}
\item Comparing the second and the last term in \eqref{CallFviaFFT} we see that the representation of the Black-Scholes part remains same as under derivation of \eqref{skewATM} but $T$ has to be replaced by $T-s$.

\item The forward characteristic function $\phi_{s,T}(u)$ doesn't depend on the stock prices $S$ and $S_s$.

\item When deriving \eqref{skewATM} one sets $k = \log (F(S,T)/K) = 0$, so here $k$ is the forward ATM log strike. In \eqref{CallFviaFFT} $k = log(K) = 0$ by the definition of $K$. Hence, despite both $k$ have a slightly different meaning, to compute the ATM implied skew (or forward skew) they both should vanish.

\item Despite \eqref{CallFviaFFT} is written using the Carr-Madan representation, it is straightforward to re-write it by using another flavor of the FFT formula \citep{Lewis:2000, Lipton2001} (as this is done in \citep{Gatheral2006}).

\end{enumerate}
With this consideration, the final result in  \eqref{skewATMForward} immediately follows.
\end{proof}
\end{proposition}

\subsection{Construction of the appropriate forward characteristic function}

In Section~\ref{example} we extended our original ADOL-Heston model by making a special choice of the function $m(t,h,H)$. This choice allows the model to catch some typical features of the vanilla skew observed for the rough volatility models. Again, to remind the asymptotic behavior of the implied ATM skew in our model at $T \to 0$ is close but not exactly same as for the rough volatility models, however, seems to be good enough to explain some stylized behavior demonstrated by the market (see the discussion in Section~\ref{disc}). And the form of $m(t,h,H)$ is critical to achieve this.

Due to similarity of representation in \eqref{skewATM} and \eqref{CallFviaFFT}, for the forward starting options an appropriate construction of the forward  characteristic function $\phi_{s,T}(u)$ is also of a paramount importance. Indeed, by using the tower rule for expectations one can write
\begin{align}
\phi_{s,T}(u) &= \EQ\left[e^{\iu u (\calS_T - \calS_s)} \;\middle|\; S, v, \calV \right] = \EQ \left[\EQ \left[e^{\iu u (\calS_T - \calS_s)} \;\middle|\; S_s, v_s, \calV_s \right] \;\middle|\; S, v, \calV \right].
\end{align}
The last expectation can also be conditioned on the variable $h$ instead of $\calV$ since by definition $h = \xi \calV - 2\sqrt{v} + \kappa(T-t)$.

The inner expectation $\EQ \left[e^{\iu u (\calS_T - \calS_s)} \;\middle|\; S_s, v_s, \calV_s \right]$ has been already computed in \eqref{ImInteg} where now the time $t$ should be replaced with $s$. Also, we use the same functional form of $m(t,h,H)$ as in \eqref{mDef}. And since in \eqref{CallFviaFFT} the forward CF should be evaluated at $\bT = T-s$, the integration limits in \eqref{ImInteg} now convert to $[t, \bT]$. At $t = s = 0$ (a vanilla option case) this model coincides with that considered in Section~\ref{CFT}.

The outer expectation on $v$ can be computed by using the explicit form of $\phi(u-\iu/2)$ in \eqref{zDef} (there it is given via the CF $z$ which is a map: $\phi(u; t,v,\calV) \mapsto z(u; t,v,h)$). To make it more transparent, first observe that using the model and the definition of $h_t$ in \eqref{ADOhestonFin}, the definition of $m(t,s,h,H)$ in \eqref{mDef} and applying \Ito lemma yields
\begin{align}
d v_t &= \zeta(h_t) t^{H-1}\ST{t > \epsilon} dt  +  \xi \nu(t) \sqrt{v_t} dW^{Q}_{2,t}, \\
d h_t &= \frac{1}{\sqrt{v_t}} \zeta(h_t)\left[ t^{H-1} - t^{H-1}\right]\ST{t > \epsilon} dt = 0. \nonumber
\end{align}
Note, that it is possible to suggest other forms of the market price of risk, not just that in \eqref{mDef}, e.g.,
\begin{align} \label{mDefF}
m(t,h,H) &= -\frac{\zeta(h)}{\xi (t+q)^{1-H}}\ST{t > \epsilon},
\end{align}
\noindent where $q$ is some constant, In this case $d h_t \ne 0$ anymore. And then a similar expression for $A_R(u;t,s,h)$ but with slightly different coefficients can be derived as well.

Since the characteristic function $\phi(u - \iu/2; s,t,v,\calV) = z(u-\iu/2; s,t,v,h)$ doesn't depend on $S_s$, it still obeys \eqref{pdeP2} with the same solution  ansatz given in \eqref{zDef} where functions $A(u; t,s,h), B(u; t,s,h)$ solve the system of ODE
\begin{align} \label{ricF}
- B'_t(u; t,s,h) &=  \frac{1}{2} \xi \rho  (1+2 \iu u) \nu(t) B(u;t,s,h)  + \frac{1}{2} \xi^2 \nu^2(t) B^2(u; t,s,h) - \frac{1}{2} (u^2 + 1/4), \\
A'_t(u;t,s,h) &=  \xi m(t,s,h,H) B(u; t,s,h), \nonumber \\
B(u;\bT,h) &= A(u;\bT,h) = 0. \nonumber
\end{align}
Since the first equation in \eqref{ricF} doesn't depend on $m(t,s,h,H)$, its solution is same as given in Section~\ref{ricSect}, but subject to a different terminal condition given in the third line of \eqref{ricF}. To confirm, again let us use asymptotic expansions. By the terminal condition in \eqref{ricF} we have $B(u;\bT,h) = 0$, and substituting it into \eqref{ric}: $B'(u;\bT,h) = \frac{1}{2} (u^2 + 1/4)$. Differentiating \eqref{ric} by $t$ and setting $t \to \bT$ we also obtain $B''(u;\bT,h) = -\frac{1}{4} \xi  \rho  (1+2 \iu u) \left(u^2+\frac{1}{4}\right) \nu(\bT)$. Overall, this gives rise to the following representation
\begin{equation} \label{seriesF}
B(u;t,s,h) = \frac{1}{2} \left(u^2+\frac{1}{4}\right) (t-\bT) \left[ 1 - \frac{1}{4} \xi  \rho  (1+2 \iu u) \nu(\bT)^2 (t-\bT)\right] + O((t-\bT)^3).
\end{equation}
This representation solves the Riccati equation in \eqref{ric} with the accuracy $O((t-\bT)^2)$. It can be seen that the leading term of the real part of $B(u;t,h)$ is $\Ree(B(u;t,h)) \propto O(\bT)$ while $\Imm(B(u;t,h)) \propto O((\bT)^2)$.

The second line of \eqref{ricF} can be explicitly integrated. When $T \to s$, (i.e. $\bT \to 0$) and $t \to 0$ the second order approximation of the final result reads
\begin{align} \label{Ares}
A_R(u;t,s,h) &= - \xi \int_0^\bT m(p,s,h,H) B_R(u;p,s,h) dp \\
&\approx -\frac{1}{4 H(2 + 3H + H^2)} \zeta(h) \left( u^2 + \frac{1}{4} \right) \left[((4 + 2H)\bT^{H+1}  + B_H^2 \rho \xi \bT^{3H + 1}\right]. \nonumber
\end{align}
Thus, $A_R(u;t,h)$ is continuous when $T \to s$. Then, as follows from \eqref{skewATMForward}, the implied skew of the forward started options blows up at $T \to s$ as $(\bT)^{-1/2}$. But as follows from the analysis for vanilla options in Section~\ref{disc}, the behavior of the implied skew at small $\bT$ is a bit more complicated.

Indeed, let us compare $A_R(u;t,h)$ in \eqref{Ares} with the similar expression in \eqref{ImInteg3} (which is recalled below) for vanilla options
\begin{align} \label{ImInteg31}
A_R(u;t,h) &= -\frac{1}{12} \zeta(h)  \left(u^2+\frac{1}{4}\right) \left[(5 - 2 H)T^{H+1} + \frac{1}{2} B^2_H \rho \xi u T^{3H+1} \right].
\end{align}
Note, that both functions depend on $H$ but in a slightly different way. However, the exponents of $T$ (for vanilla options) and $\bT$ (for forward started options) coincide. In Section~\ref{disc} we established a more accurate dependence of the implied skew by regressing the integral in \eqref{skewATMForward} and looked at the regression coefficients. In case of forward started options it can be done as well. Since we are interested in the behavior of the implied skew when $T \to 0$ (for vanilla options) and $\bT \to 0$ (for the forward started options), their asymptotic behavior should be alike, despite not exactly.

Since our ultimate goal in this paper is to judge whether Markovian approximations of rough volatility models are able to capture the behavior of the implied volatility skew at $T$ close to $t$ (for vanilla options), or $T$ close to $s$ (for the forward started options), we have already collected enough information to answer this question. The conclusion is not strict but explanatory enough.

For the {\it vanilla options} our Markovian model does capture the implied skew behavior despite producing a slightly different dependence of the skew on $H$ at $T \to t$. In particular, at very small $T$ the skew doesn't explode but rather begins to decrease. This, however, doesn't contradict to the market data as this is explained in Introduction.
For the {\it forward started options} our model demonstrates a similar behavior while the numbers in regressions are a bit different. To remind, in \citep{AlosBook,Alos2022} based on Mallavin calculus the authors claim that Markovian approximations are not able to catch the blow-up of the implied skew for the forward started options at all. As we see, nevertheless, the model catches the blow-up up to some small $T \ll 1$, but then behaves differently. Again, this doesn't contradict to the available market data on realized volatility.

Therefore, as mentioned in Introduction, we agree with \citep{AlosBook} that market data on realized volatility are not sufficient to decide which stochastic volatility model (rough or Markovian)  is more capable to reproduce the market behavior of the implied skew. Having, e.g., both the vanilla and forward implied volatilities and skews and calibrating each model to this combined set of market instruments would allow a more transparent resolution of this dilemma.

\section*{Acknowledgments}

I am grateful to Elisa Alos, Archil Gulisashvili and Igor Halperin for various useful discussions.




\end{document}